\documentstyle [12pt] {article}

\catcode`@=12
\begin{document}
\vspace{0.5in}
\oddsidemargin -.375in
\newcount\sectionnumber
\sectionnumber=0
\def\be{\begin{equation}}
\def\ee{\end{equation}}
\begin{flushright} UH-511-825-95\\April 1995\
\end{flushright}
\vspace {.5in}
\begin{center}
{\Large\bf  Semi-Leptonic  Decays of
 ${\bf \Lambda_c}$ and ${\bf \Lambda_b}$ Baryons involving Heavy to Light
Transitions and the determination of ${\bf |V_{ub}|}$  \\}
\vspace{.5in}
{\bf Alakabha Datta \\}
\vspace{.1in}
 {\it
Physics Department, University of Hawaii at Manoa, 2505 Correa
Road, Honolulu, HI 96822, USA.}\\
\vskip .5in
\end{center}

\vskip .1in
\begin{abstract}

We calculate the semi-leptonic decays of $\Lambda_c$ and
$\Lambda_b$
 baryons which involve  transitions of a heavy quark to a light quark
. In a previous work we used data on the Cabibbo favoured non-leptonic decays
of charmed
baryons $\Lambda_c \rightarrow \Lambda \pi^{+}$ and
 $\Lambda_c \rightarrow \Sigma^{+} \pi^{0 }$
 to obtain information on the form factors in the $c \rightarrow s $
transition. This form factor information is used to study the
semi-leptonic decay $\Lambda_c \rightarrow \Lambda l \nu $. Using
SU(3) symmetry and HQET it is shown how a knowledge of form factors in
$\Lambda_c \rightarrow \Lambda l \nu $ can be applied to
 $\Lambda_b \rightarrow p l \nu $
 decay and used to measure $|V_{ub}/V_{cs}|^{2}$.

\end{abstract}
\vskip .25in
\section{\bf Introduction}
Recently there has been a measurement of the form factors in the
semi-leptonic decay $\Lambda_c \rightarrow \Lambda e^{+} \nu_{e} $
\cite{cleo}, where
 a fit to the data was done using the model of K\"orner and Kramer
\cite{KK}. The model, constructed within the framework of HQET, treats
both the charm quark and the strange quark as heavy and considers $1/m_s$
expansion keeping $m_c\rightarrow\infty$. Even though the model has only
one input parameter, use of $1/m_s$ expansion is
questionable and the omission of $1/m_c$ corrections is also not justified.
For a
heavy to light transition, for example of the type
$ \Lambda_c \rightarrow \Lambda $,
the
use of HQET in the limit $m_Q \rightarrow \infty $ allows one to express all
the form factors in terms of only two form factors \cite{mR}. Semileptonic
decays of $\Lambda_c $ have been studied in this limit \cite{Grin,KK}.
 In
Ref. \cite{Al,Al1} we constructed a model for the form-factors which takes
into account $1/m_Q$ corrections for heavy-to-light transitions
 of the $\Lambda$ type baryon (light degrees in
spin 0 state). In this work we use the form factors
calculated in Ref. \cite{Al1} to study the decays $\Lambda_c \rightarrow
\Lambda l \nu $ and  $\Lambda_b \rightarrow p l \nu
$.

The paper is organized in the following way. In the next section we present
the formulae
 for the calculation of asymmetries and decay rates for the
 semi-leptonic decays
$\Lambda_c \rightarrow \Lambda l \nu $ and $\Lambda_b \rightarrow p l \nu $
while in the third section we present our results.

\section{\bf Formalism}

Here we present the formalism for the semi-leptonic decays of
$\Lambda_c \rightarrow \Lambda l \nu $ and $\Lambda_b \rightarrow p l \nu $.
 The hadronic part of the amplitude is the matrix elements of the
 weak quark current between baryonic states
that is
parametrized in terms of form factors. We define the six vector and axial
vector form factors through the following equations
\begin{eqnarray}
 \left < B'(p',s')
\mid \bar{q} \ \gamma^\mu \ Q \mid B_Q (p, s)
\right >
 =   \bar{u}_{B'} (p', s')
\left [f_1 \gamma^\mu - i \frac{f_2}{m_{B_c}}\sigma^{\mu\nu}
 q_\nu + \frac{f_3}{m_{B_c}} q^\mu \right ] u{_B{_Q}} (p, s)
\nonumber \\
 \left < B'(p', s')
\mid \bar{q} \ \gamma^\mu \gamma^5 \ Q \mid B_Q (p,s)
\right >
  =   \bar{u}_{B'} (p', s')
\left [g_1 \gamma^\mu - i \frac{g_2}{m_{B_Q}}\sigma^{\mu\nu}
 q_\nu + \frac{g_3}{m_{B_c}} q^\mu \right ]\gamma^5 u{_B{_Q}}(p, s)\
\end{eqnarray}
where $q^\mu = p^\mu-p'^\mu$ is the four momentum transfer, $B_Q$ is the
baryon with a heavy quark and $B'$ is the light baryon.
We have shown in Ref. \cite{Al,Al1} that we can write the form factors
in the above equation
in terms
of two form factors $F_1$ and $F_2$ provided some assumptions are made
about the $1/m_c$ corrections.
 For both $\Lambda_c$ and
 $\Lambda_b$ semi-leptonic decays  we will work in the approximation that the
lepton mass is negligible. The decay $\Lambda_c\rightarrow\Lambda_s
l\nu_{l}$ proceeds via $\Lambda_c\rightarrow\Lambda_s
W_{off-shell}$ followed by $W_{off-shell}\rightarrow l\nu_l$. Following
\cite{KK} we define the  helicity amplitudes
which are given  by
\begin{eqnarray}
H_{\lambda_2,\lambda_{W}} &  = &  H^V_{\lambda_2,\lambda_{W}}+
 H^A_{\lambda_2,\lambda_{W}}\nonumber\\
H^{V,(A)}_{-\lambda_2,-\lambda_{W}} &= &+ (-) H^{V,(A)}_{\lambda_2,\lambda_{W}}
\end{eqnarray}
where $\lambda_2,\lambda_{W}$ are the polarizations of the daughter baryon and
the W-boson respectively. In terms
of the form factors the helicity amplitudes are given by
\begin{eqnarray}
H^{V}_{1/2,0} & = & a_{-}\left [(M_1+M_2)f_1+\frac{q^2}{m_{B_c}}f_2\right
]V(q^2)
\nonumber\\
H^{A}_{1/2,0} &=& a_{+}\left [-(M_1-M_2)g_1+\frac{q^2}{m_{B_c}}g_2\right
]A(q^2)
\nonumber\\
H^{V}_{1/2,1}& = & \sqrt{2Q_{-}}\left [ -f_1 - \frac{(M_1+M_2)}{m_{B_{c}}}
f_2 \right ]V(q^2)
\nonumber\\
H^{A}_{1/2,1} & = & \sqrt{2Q_{+}}\left [ g_1 - \frac{(M_1-M_2)}{m_{B_c}}g_2
\right ]A(q^2)
\end{eqnarray}
where
\begin{eqnarray}
Q_{\pm}& =& {(M_{1} \pm  M_2)}^2 - q^2 \quad ; \quad a_{\pm} =
\sqrt{\frac{Q_{\pm}}{q^2}}\nonumber\\
V(q^2) & =& \frac{(1-\frac{q^2_{max}}{{m^V_{FF}}^2})^2}
{(1-\frac{q^2}{{m^V_{FF}}^2})^2} \quad ; \quad
A(q^2) = \frac{(1-\frac{q^2_{max}}{{m^A_{FF}}^2})^2}
{(1-\frac{q^2}{{m^A_{FF}}^2})^2}
\end{eqnarray}
where $M_1,M_2$
are the parent and daughter baryon masses and
 $m^{V,A}_{FF}$ are the appropriate pole masses. The decay
  $\Lambda_c\rightarrow\Lambda_s
l\nu_{l}$ is analysed by looking at the two sided cascade decay
 $\Lambda_c\rightarrow
\Lambda_s[\rightarrow p\pi^{-}]+
W[\rightarrow l \nu_l]$. The
 normalized four-fold joint angular distribution for the decay
$\Lambda_1\rightarrow
\Lambda_2[\rightarrow a({\frac{1}{2}}^{+})+b(0^{-}]+
W[\rightarrow l \nu_l]$
 is given, following
   the notation in Ref. \cite{KK}, as
\begin{eqnarray}
\frac{d\Gamma}{dq^{2}d\chi d\cos\Theta d\cos\Theta_{\Lambda}} &=&
B(\Lambda_2\rightarrow
a+b)\frac{1}{2}\frac{G^2}{{(2\pi)}^4}\frac{q^2p}{24M_1}|V_{Q_1,Q_2}|^2
\left[T_1+T_2+T_3\right]\
\end{eqnarray}
where
\begin{eqnarray*}
 T_1 &= &\left[\frac{3}{8}{(1\pm\cos\Theta)}^2|H_{1/2 \, 1}|^2
(1+\alpha_{\Lambda}\cos\Theta_{\Lambda})
 +  \frac{3}{8}
{(1\mp\cos\Theta)}^2|H_{-1/2 \, -1}|^2
(1-\alpha_{\Lambda}\cos\Theta_{\Lambda})\right]\nonumber\\
 T_2 &= & \frac{3}{4}\sin^2\Theta\left[|H_{1/2 \, 0}|^2
(1+\alpha_{\Lambda}\cos\Theta_{\Lambda})
  +  |H_{-1/2 \, 0}|^2
(1-\alpha_{\Lambda}\cos\Theta_{\Lambda})\right]\nonumber\\
T_3 &= &
\mp\frac{3}{2\sqrt{2}}\alpha_{\Lambda}\cos\chi\sin\Theta\sin\Theta_{\Lambda}
\left[(1\pm\cos\Theta)Re(H_{-1/2 \, 0}H^*_{1/2 \, 1}) +
(1\mp\cos\Theta)Re(H_{-1/2 \, 0}H^*_{-1/2 \, -1})\right] \
\end{eqnarray*}
where $p=\sqrt{Q_{+}Q_{-}}/2M_1$, the upper and lower signs in the above
equation hold for the $l^{-}\nu_l$ and $l^{+}\nu_l$ leptonic final states
respectively
and $V_{Q_1,Q_2}$ is the CKM element for the $Q_1\rightarrow Q_2$ transition.
The polar angles are defined in Ref. \cite{KK}. By integrating over two of the
angles one can look at the following distributions
\begin{eqnarray}
\frac{d\Gamma}{dq^{2}d\cos\Theta_{\Lambda}} & \propto & 1+\alpha
\alpha_{\Lambda}
\cos\Theta_{\Lambda}\nonumber\\
\frac{d\Gamma}{dq^{2} d\cos\Theta} & \propto & 1\pm 2 \alpha^{\prime}\cos\Theta
+\alpha^{\prime\prime}{\cos\Theta}^{2}\nonumber\\
\frac{d\Gamma}{dq^{2}d\chi} & \propto & 1\mp \frac{3{\pi}^2}{32\sqrt{2}}\gamma
\alpha_{\Lambda} \cos\chi \
\end{eqnarray}
The first distribution in the above equation gives the polar angle
distribution for the cascade decay $\Lambda_s\rightarrow p \pi^{-}$. The
second distribution is the polar angle distribution for the deacy
$W \rightarrow l \nu_l$, while the third distribution is the azimuthal angle
distribution.
 The asymmetries are given by the expressions below
 \cite {KK}
  and will depend only
on the ratio $F^0_2(\omega=1)/F^0_1(\omega=1) $ in our model for the
form factors
\begin{eqnarray}
\alpha &=& \frac{|H_{1/2 \ 1}|^2 - |H_{-1/2 \ -1}|^2 + |H_{1/2 \ 0}|^2
- |H_{-1/2 \ 0}|^2}{|H_{1/2 \ 1}|^2 + |H_{-1/2 \ -1}|^2 + |H_{1/2 \ 0}|^2
+ |H_{-1/2 \ 0}|^2} \nonumber\\
\alpha^{\prime} &=& \frac{|H_{1/2 \ 1}|^2 - |H_{-1/2 \ -1}|^2 }
{|H_{1/2 \ 1}|^2 + |H_{-1/2 \ -1}|^2 +2 \ (|H_{1/2 \ 0}|^2
+ |H_{-1/2 \ 0}|^2)}\nonumber\\
\alpha^{\prime \prime} &=& \frac
{|H_{1/2 \ 1}|^2 + |H_{-1/2 \ -1}|^2 - 2 \ (|H_{1/2 \ 0}|^2
+ |H_{-1/2 \ 0}|^2)}
{|H_{1/2 \ 1}|^2 + |H_{-1/2 \ -1}|^2 +2 \ (|H_{1/2 \ 0}|^2
+ |H_{-1/2 \ 0}|^2)} \nonumber\\
\gamma &=& \frac{2 \ Re  (H_{-1/2 \ 0}H^*_{1/2 \ 1} +
 H_{1/2 \ 0}H^*_{-1/2 \ -1})}
{|H_{1/2 \ 1}|^2 + |H_{-1/2 \ -1}|^2 + |H_{1/2 \ 0}|^2
+ |H_{-1/2 \ 0}|^2}\
\end{eqnarray}
 for unpolarized $\Lambda_c$. For polarized $\Lambda_c$ one  has similar
decay distributions and we refer the reader to Ref. \cite{KK} for the relevant
details. We give here the expression for the asymmetries
\begin{eqnarray}
\alpha_{P} &=& \frac{|H_{1/2 \ 1}|^2 - |H_{-1/2 \ -1}|^2 - |H_{1/2 \ 0}|^2
+ |H_{-1/2 \ 0}|^2}{|H_{1/2 \ 1}|^2 + |H_{-1/2 \ -1}|^2 + |H_{1/2 \ 0}|^2
+ |H_{-1/2 \ 0}|^2} \nonumber\\
\gamma_{P} &=& \frac{2 \ Re (H_{1/2 \ 0}H^*_{-1/2 \ 0})}
{|H_{1/2 \ 1}|^2 + |H_{-1/2 \ -1}|^2 + |H_{1/2 \ 0}|^2
+ |H_{-1/2 \ 0}|^2}\
\end{eqnarray}
All the asymmetries have limiting values as $q^2\rightarrow
{q^{2}}_{max}$. All the polar asymmetries vanish at this limit while the
azimuthal asymmetryies $\gamma\rightarrow 2\sqrt{2}/3$ and
$\gamma_P\rightarrow -1/3$. At the $q^2=0$ point however the limiting
value of the asymmetries
$\alpha$, $\alpha_P$ and $\gamma_P$ depend on the dynamics of the
semi-leptonic decay
and we will comment on this issue further in the next section. The other
asymmetries $\alpha^{\prime}$ and $\gamma$ tends to zero while
$\alpha^{\prime\prime}$ tend to $-1$ in the $q^2=0$ point.
These asymmetries were studied for different representative values of
the ratio $r_{KK}=f_2/f_1$ \cite{KK} which is related to the
 ratio $r=F^0_2(\omega=1)/F^0_1(\omega=1) $ via $r_{KK}=r/(2+r)$.  The
fit, performed by
Ref. \cite{cleo},
to the semileptonic decay data on $\Lambda_c\rightarrow
 \Lambda_s e^{+}\nu_{e}$ uses the Korner Kramer(KK) model to
extract $r_{KK}$ and the asymmetry $\alpha$ \cite{cleo}. Since, as already
mentioned in the introduction, the KK model
does not include $1/m_c$ corrections and the use of $1/m_s$ expansion is
questionable, a more correct approach would be therefore to do a fit to the
semi-leptonic data including $1/m_c$ corrections to extract the ratio
$r(r_{KK})$ and hence the asymmetries. The formula for the calculation of
the absolute decay rate for the semi-leptonic decay of baryons is given in
Ref. \cite{DR}

\section{\bf Results}

In our analysis we have studied the semi-leptonic decays
for $|r|\leq 1$ and $0 \leq F_1^{0}(\omega=1)\leq 1$. In Ref. \cite{Al1}
we found the $1/m_c$ expansion to be valid for
$|r| \leq 1$ and from the study of non-leptonic two body charmed baryon decays
$ F_1^{0}(\omega=1) \leq 1$ was obtained ( we assumed
 $F_1^{0}(\omega=1) >0$ ). In our analysis we found that the $1/m_Q$
corrections depend on the value of $r$. As far as the $1/m_Q$ corrections to
the form factors are concerned for the $\Lambda_c$ decay the $1/m_c$
corrections to the form factors are $<50\%$ for $-0.62\leq r \leq1$. For
$\Lambda_b$ decays the $1/m_b$ corrections are small and are less than $30 \% $
for $-0.48\leq r \leq 1$.
 Fig.1  shows $\alpha(q^{2})$ versus $q^{2}$ for different $ r$ values. We
note that for $ -0.5\leq r \leq 1$ the asymmetry $\alpha$ is not very
sensitive to $r$. This feature is common to most of the calculated integrated
asymmetries. In Fig.2 we show $\alpha(q^{2}=0)$ versus $r$. In the KK
model or in general with $m_c \rightarrow\infty$, $\alpha(q^{2}=0)=-1$ for
all $r$. Fig.2 shows the effect of including $1/m_c$ corrections.  Fig.3 and
Fig.4 show the predictions for the integrated  asymmetries for different $r$
for unpolarised and polarised
 $\Lambda_c$ and except for $\gamma$ we see that the integrated
asymmetries are insensitive to r in the range $ -0.5\leq r \leq 1$.
The $1/m_c$ corrections to the integrated asymmetries again depend on
the value of r and can be as large as $40 \%$.

The decay
rates depend on both $r$ and $F_1^{0}(\omega=1)$. So instead of calculating
the individual decay rates for $\Lambda_c \rightarrow \Lambda l \nu $ and
$\Lambda_b \rightarrow p l \nu $, to reduce the uncertainties from these
sources
 we calculate the ratio of the decay rates
which will depend only on $r$ as the factor $F_1^{0}(\omega=1)$ cancels.
Fig.5 shows this ratio as a function of $r$ with and without $1/m_Q$
corrections with a monopole and dipole form for the form factors. We see that
the $1/m_Q$ corrections are small for the dipole form factors. The ratio
above will
 receive corrections of the order
 $\sim 1/m_{Q}^{2}$ and higher where $ m_{Q} $ is the c or b
quark. An estimate of these corrections will depend on the value of $r$
which can be extracted by performing a fit to the
$\Lambda_c \rightarrow \Lambda e^{+} \nu_{e} $ data.
 The final result can be written
as
\begin{eqnarray}
\frac{\Gamma(\Lambda_b \rightarrow p l \nu)}
 {\Gamma(\Lambda_c \rightarrow \Lambda l \nu)}=R_0(r)[1+\Delta_1(r)+
\Delta_2(r) + ......]|\frac{V_{ub}}{V_{cs}}|^{2}\
\end{eqnarray}
where $R_0$ is the ratio with $m_Q \rightarrow \infty$ and $\Delta_{j}(r)$
represents the corrections to the ratio due to
 $ 1/m_{Q}^{j}$ corrections to the form factors.
 Since we do not know $r$ we can make
an estimate of the ratio above by using the value of $r$ extracted from
non-leptonic charmed baryon decays with dipole form factors
 \cite{Al1} i.e $r=r_0=-0.47(r_{KK}=-0.31)$.
 We write
\begin{eqnarray}
r &=& r_0 \pm \Delta r \nonumber\\
R_0(r) &\approx & R_0(r_0) \pm \Delta r R'_0(r_0)\
\end{eqnarray}
where $R'_0(r_0)$ is the derivative of $R_0(r)$ at $r=r_0=-0.47$, and where
$\Delta r$ may also include experimental uncertainties in extracting $r$. So
an estimate of the ratio defined in eqn.(9) can be made through the
following approximation
\begin{eqnarray}
\frac{\Gamma(\Lambda_b \rightarrow p l \nu)}
 {\Gamma(\Lambda_c \rightarrow \Lambda l \nu)}\approx R_0(r_0)[1
\pm \Delta r R'_0(r_0)/R_0(r_0) +
\Delta_1(r_0)+
\Delta_2(r_0) + ......]|\frac{V_{ub}}{V_{cs}}|^{2}\
\end{eqnarray}
For monopole and dipole form factors we then have
\begin{eqnarray}
\frac{\Gamma(\Lambda_b \rightarrow p l \nu)}
 {\Gamma(\Lambda_c \rightarrow \Lambda l \nu)}\approx 196.65[1
\pm 0.78\Delta r  +
0.02+
\Delta_2(r_0) + ......]|\frac{V_{ub}}{V_{cs}}|^{2}\
\end{eqnarray}
and
\begin{eqnarray}
\frac{\Gamma(\Lambda_b \rightarrow p l \nu)}
 {\Gamma(\Lambda_c \rightarrow \Lambda l \nu)}\approx 83.41[1
\pm 0.94\Delta r  -
0.15+
\Delta_2(r_0) + ......]|\frac{V_{ub}}{V_{cs}}|^{2}\
\end{eqnarray}
Naively we can set $ \Delta_2(r_0)\approx \Delta_1(r_0)/2m_Q$ and so we see
that $\Delta_2(r_0)\stackrel{>}{\sim} .05$ for the dipole form factors and
$\Delta_2(r_0)$ is negligible for the monopole form factors.
Measurement of the ratio $R$ can be used to extract $|{V_{ub}}|/|{V_{cs}}|$.
Clearly the largest uncertainty in $R$ comes from the assumed $q^{2}$
dependence of the form factors and to reliably extract $|{V_{ub}}|/|{V_{cs}}|$
using $R$ the $q^{2}$
dependence of the form factors must be measured. Our model
for the form factors at $\omega=1$ can be
tested by studying the decay distribution in $\Lambda_c \rightarrow \Lambda
 l \nu $. With enough statistical power such a study could  determine the
$q^{2}$ dependence of the form factors. So if our model
for the form factors at $\omega=1$ is tested successfully
by doing a fit to the $\Lambda_c \rightarrow \Lambda
 l \nu $ decay distribution, and if $r$ and information on the $q^{2}$
dependence of form factors extracted, we can use the ratio in eqn.(9) to
determine $|{V_{ub}}|/|{V_{cs}}|$ with a small error.

It is interesting to compare this method of extracting $|{V_{ub}}|/|{V_{cs}}|$
and hence $|{V_{ub}}|$ with other methods
 of extracting $|{V_{ub}}|$. One method is to use semi-leptonic inclusive $B$
decays and study the decay distribution near the end point of the spectrum in
the narrow window $2.3 \leq E_l \leq 2.6 $ GeV \cite {BM}. However in this
region the $b \rightarrow u $ transition is subject to large QCD radiative
corrections as well as large non-perturbative corrections which are entangled
with each other and theoretical calculations are model dependent.
Another way to extract $|{V_{ub}}|$ is through the study of exclusive $B$
decays \cite{Kim} and this approach is similar to our approach with the
difference that we are working with baryons instead of mesons. The
calculations
of form factors in exclusive decays of $B$ mesons are done in specific
models, like the quark model, or with chiral
 Lagrangian with heavy meson. In Ref. \cite{Burd} it is
suggested that $|V_{ub}|$ may be extracted from the ratio
of the lepton distribution in $ B \rightarrow \pi l \nu $ and
$ D \rightarrow \pi l \nu $ with an theoretical uncertainty of 10-20 $\%$. The
relevant form factors are calculated up to order $1/m_Q$ in the soft pion limit
with the additional assumption that only intermediate states degenerate with
the ground state meson contribute and so the calculations of the form
factors are model dependent. In Ref. \cite{Dib}
the ratio
of the lepton distribution in $ B \rightarrow \rho l \nu $ and
$ D \rightarrow \rho l \nu $ at maximum $q^{2}$ is used for an extraction of
$|V_{ub}|$. The $1/m_Q$ corrections are calculated in quark model and the
theoretical uncertainty in the measurement of $|V_{ub}|$ is small. However
at maximum $q^{2}$ the decay rates vanish, actual measurements are
performed away from maximum $q^{2}$ where additional form factors contribute
adding to the theoretical uncertainty in the measurement of $|V_{ub}|$.
Another approach to measuring $|V_{ub}|$, which is the mesonic counterpart of
our method, is given in Ref. (\cite{Du}) where experimental data on $D
\rightarrow {\bar K}$ semi-leptonic decay are used to calculate the
$ B \rightarrow \pi$ semileptonic decay. The $q^{2}$ dependence of the form
factors is assumed to be of the monopole type, no $1/m_Q$
corrections are considered and the extraction of $|V_{ub}|$ is found to
 depend on the
value of $f_{D_s}$ and $f_{B}$ whose values are not known very accurately.
Since most approaches to extracting $|V_{ub}|$ from $B$ meson decays are
model dependent and can involve significant theoretical
 uncertainties, extraction of
$|V_{ub}|$ using heavy baryon decays is interesting as it is possible that
theoretical uncertainties in heavy baryon decays might be small.

The predictions for the decay rates for
$ \Gamma(\Lambda_c \rightarrow \Lambda l \nu)$ and
$\Gamma(\Lambda_b \rightarrow p l \nu)$ are subject to significant $1/m_Q$
corrections and may not be reliable. Nevertheless we give the predictions for
the decay rates with $r=-0.47$ and $F^{0}(\omega=1)=0.46$\cite {Al1}
\begin{eqnarray}
\Gamma(\Lambda_c \rightarrow \Lambda l \nu) &= &
5.36\times 10^{10} s^{-1}\\
\Gamma(\Lambda_b \rightarrow p l \nu) &=& |V_{ub}|^{2}
6.48\times 10^{12} s^{-1}\
\end{eqnarray}
The Particle Data Group gives $\Gamma(\Lambda_c \rightarrow \Lambda e^{+} X)
=(7.0 \pm 2.5)\times 10^{10} s^{-1}$. Measurement
of $\Gamma(\Lambda_b \rightarrow p l \nu) $ is likely to be made at LEP in
the future.
 The numbers for the individual decay
rates are calculated using
 $|V_{cs}|=0.9745$ \cite{PDG}, the
  pole masses $m_V \cong 5.37$ GeV  and $m_A \cong 5.80$ GeV\cite{isgur}
 and the
quark masses $m_b=
4.74$ GeV and $m_u=
0.005$ GeV\cite{QM}

In conclusion we have studied the semi-leptonic decays of
charmed and bottom baryons involving transition of a heavy to light quark
based on a model for form factors that includes $1/m_Q$ corrections. We also
suggest a method for determining $|V_{ub}|/|V_{cs}|$.

{\bf Acknowledgement } :
I would like to thank Professor Sandip Pakvasa and
 Professor Tom Browder for useful discussions. This
 work was supported in part by US
D.O.E grant \# DE-FG 03-94ER40833.

\subsection{Figure Captions}
\begin{itemize}
\item[{\bf Fig. 1}] The asymmetry $\alpha(q^{2})$ versus $q^{2}$ for
some representative values of the form factor ratio $r$.

\item[{\bf Fig. 2}] The asymmetry $\alpha(q^{2}=0)$ versus $r$. The
$m_c \rightarrow \infty$ prediction for $\alpha(q^{2}=0)$ is $-1$ for all $r$.
This figure therefore shows the effect of $1/m_c$ corrections.

\item[{\bf Fig. 3}] The average asymmetries
as functions of $r$ when the initial heavy baryon is unpolarised.

\item[{\bf Fig. 4}] The average asymmetries
as functions of $r$ when the initial heavy baryon is polarised.

\item[{\bf Fig. 5}] The ratio of the decay rates of
${\Gamma(\Lambda_b \rightarrow p l \nu)}/
 {\Gamma(\Lambda_c \rightarrow \Lambda l \nu)}$ versus $r$. $R_0$ and $R$
represents the ratio without and with $1/m_Q$ corrections. Monopole and
dipole form for the form factors are used for the $q^{2}$ dependence of the
form factors.
\end{itemize}
\end{document}